\documentstyle[11pt,newpasp,epsf,twoside]{article}
\def\edcomment#1{\iffalse\marginpar{\raggedright\sl#1\/}\else\relax\fi}
\reversemarginpar

\begin{document}
\title{The Power of Exploratory \textit{Chandra} Observations}
\author{Sarah C. Gallagher}
\affil{University of California, Los Angeles, Division of Astronomy \& Astrophysics, 405 Hilgard Avenue, Los Angeles, CA 90095}
\author{Gordon T. Richards}
\affil{Princeton University Observatory, Peyton Hall, Princeton, NJ 08544}
\author{W. Nielsen Brandt \& George Chartas}
\affil{The Pennsylvania State University, Department of Astronomy \& Astrophysics, 525 Davey Laboratory, University Park, PA 16802}

\begin{abstract}
With its excellent spatial resolution, low background, and hard-band
response, the {\em Chandra} X-ray Observatory is ideal for performing
exploratory surveys. These efficient, sensitive observations can place
constraints on fundamental properties of a quasar continuum including the
X-ray luminosity, the ratio of X-ray to UV power, and the X-ray
spectral shape. To demonstrate the power of such surveys to provide
significant insight, we consider two examples,
a Large Bright Quasar Survey sample of broad absorption line quasars and
a sample of Sloan Digital Sky Survey (SDSS) quasars with extreme C~{\sc iv} blueshifts.  
In both cases, exploratory {\em Chandra} observations provide important
information for a physical understanding of UV spectroscopic
differences in quasars.

\end{abstract}

\vspace{-0.3cm}

\section{Introduction}

X-ray emission appears to be a universal signature of quasar spectral energy
distributions, confirming expectations from accretion physics.
Based on rapid variability of soft X-rays
in conjunction with the standard black-hole paradigm, these
photons are believed to be emitted from the region immediately
surrounding the black hole.
Energetically, X-rays are significant, contributing 2--20$\%$
of the bolometric luminosity.  X-ray observations are thus an
important component of any multi-wavelength campaign to probe quasar
populations.

The excellent spatial resolution of the {\em Chandra} High
Resolution Mirror Assembly and the effective background rejection of
the ACIS instrument make this combination uniquely powerful for 
quasar surveys.  
For reference, during a 5~ks observation, the
0.5--8.0~keV background within a $2\arcsec$-radius source region is
typically $\sim$0.1~ct. Because ACIS is photon-limited even beyond 
$100$~ks (Alexander et al.\ 2003), the point-source detection limit scales {\em
linearly} with exposure time, unlike the $\sqrt{t} $ dependence common in other
wavelength bands. In conjunction with sub-arcsec positional
accuracy, known optical point sources can be robustly detected with 3--5
photons. In 5 ks, this corresponds to a \mbox{0.5--8.0~keV} flux of 
$\sim7\times10^{-15}$~erg~cm$^{-2}$~s$^{-1}$ for a typical quasar
X-ray spectrum.

A 3--7~ks {\em Chandra} exposure, the regime of exploratory
observations, is generally insufficient for gathering enough X-rays
for spectral analysis of a quasar.  However, the strategy of exploratory
observations enables the extension of results from spectroscopic
observations of individual
targets to larger, well-defined samples, and the investigation of
connections between X-ray properties and other wavelength regimes.
From these datasets, standard X-ray
observables are \mbox{0.5--8.0~keV} flux, hardness ratio,\footnote{The
  hardness ratio is defined to be $(h-s)/f$, where $h$=2--8~keV ct,
  $s$=0.5--2.0~keV ct, and $f$=0.5--8.0~keV ct.} and $\alpha_{\rm
  ox}$.\footnote{The quantity $\alpha_{\rm ox}$ equals
  $0.384\log(f_{\rm X}/f_{\rm 2500})$
 where $f_{\rm X}$ and $f_{\rm 2500}$ are the flux densities at
 rest-frame 2~keV and 2500~\AA, respectively.}

We briefly describe the initial results from two exploratory {\em Chandra} quasar
surveys to illustrate the utility of this observing strategy.  Other
examples in the literature of successful applications of this approach to quasar studies 
include surveys of high-$z$ (e.g., Brandt et al.\ 2002; Vignali et
al.\ 2003), red (Wilkes et al. 2002), and X-ray weak (Risaliti et
al.\ 2003) quasars. 

%%%%%%%%%%%%%%%%%%%%%%%%%%%%%%%%%%%%%%%%%%%%%%%%%%%%%%%%%%%%%%%%%%%%
\begin{figure}
\plotone{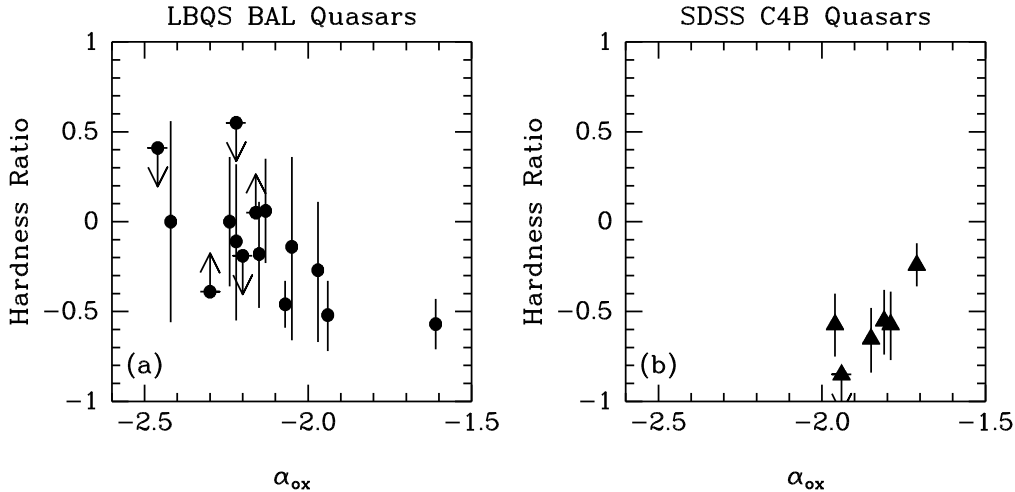}
\caption{Hardness ratios versus $\alpha_{\rm ox}$ from exploratory
  {\em Chandra} surveys for two quasar samples.
($a$) The Large Bright Quasar Survey broad absorption line quasar
  sample (Gallagher et al.\ 2003). ($b$) The SDSS C~{\sc iv}
  blueshift sample (Richards et al., in prep.).}
\end{figure}
%%%%%%%%%%%%%%%%%%%%%%%%%%%%%%%%%%%%%%%%%%%%%%%%%%%%%%%%%%%%%%%%%%%%

\vspace{-0.1cm}

\section{X-ray Insights from the LBQS BAL~Quasar {\em Chandra} Survey}

We are in the process of performing the largest exploratory survey to
date of a well-defined sample of broad absorption line (BAL) quasars
drawn from the Large Bright Quasar Survey (LBQS).  Since BAL~quasars are
known to be very faint X-ray sources (e.g., Green \& Mathur
1996; Gallagher et al.\ 1999), exploratory observations are the only
means of observing sufficient numbers to
determine the X-ray properties of the population as a whole.
The {\em Chandra} data alone
are revealing. As seen in Figure~1a, the hardness ratio appears to be  
anti-correlated with $\alpha_{\rm ox}$.  This indicates that the
X-ray weakest BAL~quasars have the hardest spectra, consistent with
the understanding from spectroscopic observations of a handful of
objects (e.g., Gallagher et al. 2002) that the
X-ray spectra are heavily absorbed. Examining the connection between
the X-ray and UV absorption properties of the quasars has also placed
observational contraints on quasar disk-wind models (Gallagher et
al. 2003).

In addition to exploratory {\em Chandra} observations, this
sample is also being targeted by both SCUBA (PI Priddey) and {\em
 SIRTF} to characterize the submm through hard X-ray spectral energy
distributions of BAL~quasars as a whole. 

\vspace{-0.1cm}

\section{The Connection Between C~{\sc iv} Blueshift and X-ray Properties}

High-ionization broad emission lines such as C~{\sc iv} have been known to
yield redshifts systematically lower than those measured from Mg~{\sc ii}
(e.g., Tytler \& Fan 1992); i.e., these C~{\sc iv} lines are
blueshifted relative to the systemic velocity.  In a study of $\sim
800$ SDSS quasars with 
$1.5 \le z \le 2.2$, Richards et al.\ (2002) found that the C~{\sc iv}--Mg~{\sc
  ii} velocity shifts ranges over $\ge2000$~km~s$^{-1}$.
Furthermore, the C~{\sc iv} blueshift (hereafter C4B) was correlated
with UV properties, most notably the relative $\Delta(g-i)$ color
(Richards et al.\ 2003).  That is, the bluest
quasars typically exhibited the largest C4Bs.  Positing that the C4B
might result from the orientation of the accretion disk or the
opening angle of the disk wind, we proposed an exploratory {\em
  Chandra} survey to investigate this hypothesis.

%%%%%%%%%%%%%%%%%%%%%%%%%%%%%%%%%%%%%%%%%%%%%%%%%%%%%%%%%%%%%%%%%%%%
\begin{figure}
\plotone{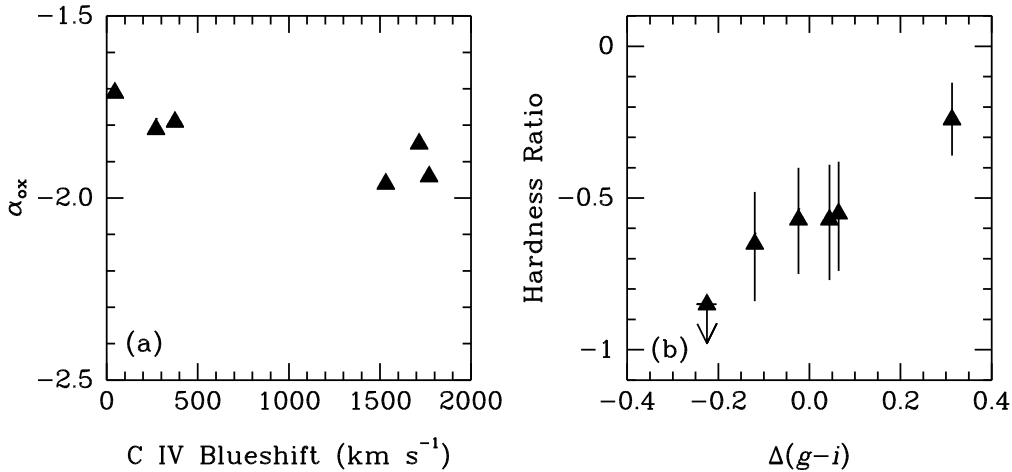}
\caption{ Measured X-ray properties from the C4B {\em Chandra} survey
  versus UV properties.
($a$) $\alpha_{\rm ox}$ versus C~{\sc iv} blueshift.
($b$) Hardness ratio versus $\Delta(g-i)$ color. Redder colors are to the right.}
\end{figure}
%%%%%%%%%%%%%%%%%%%%%%%%%%%%%%%%%%%%%%%%%%%%%%%%%%%%%%%%%%%%%%%%%%%%

Six targets, three each from the extreme ends of the C4B distribution,
were approved for Cycle 4 observations.  
While any trends based on six data points need verification, the
initial results from this small survey are intriguing.  
Figure~1b shows that hardness appears to increase with
$\alpha_{\rm ox}$ for the SDSS C4B quasar sample.  
Though this trend is not statistically significant (Spearman's
rank-order correlation coefficient, $r_{s}$, is 0.67 for a significance
level, $p_{rs}$, of 0.15\footnote{The significance
  level ranges from 0.0--1.0 with a small value
  indicating a significant correlation.}), the fact that the hardest sources
are not X-ray weaker is relevant. 
%That is, the weaker X-ray sources 
%have softer spectra, exactly the opposite of what is seen for the
%LBQS~BAL~quasar sample.  
This suggests that the hardness of the  spectra may not be due 
to intrinsic absorption in the same way as we see with the BAL~quasars.  
Extending the study to the UV properties, we tested $\alpha_{\rm
  ox}$ versus C4B and hardness ratio versus  $\Delta(g-i)$
(see Figure~2).  Though $\alpha_{\rm ox}$ and C4B are consistent
with being uncorrelated ($r_{s}$=$-$0.77, $p_{rs}$=0.07), the hardness
ratio is significantly correlated with $\Delta(g-i)$ ($r_{s}$=$-$0.99, $p_{rs}=3\times10^{-4}$).
As shown in Figure~2b, the bluer quasars appear to have softer X-ray
spectra, i.e., more negative values of the hardness ratio.
This is in line with expectation if UV continuum
color is solely related to intrinsic obscuration.  However, the
lack of connection with $\alpha_{\sc ox}$ makes this interpretation uncertain.
The connection of UV spectroscopic properties to X-ray emission in these objects 
implies a physical connection, and more data to investigate
this claim are certainly warranted.

This experiment also illustrates the value of the SDSS to
multiwavelength quasar studies.  Given the large
number and uniform data quality of the available SDSS quasars, samples
can be chosen with precision. 
Since hardness ratio can vary with $z$ due to absorption % and Galactic column
%density, 
and $\alpha_{\rm ox}$ is a function of $l_{\rm 2500}$, the
luminosity density at rest-frame 2500~\AA\ (Vignali et al. 2003), sample
tuning significantly reduces potential selection biases.
In this C4B quasar survey, the redshifts range from
$z$=1.65--1.89 and $l_{\rm 2500}$ spans only a factor of $\sim5$.
%, and the Galactic column densities are all within
%$20\%$ of the mean.  
The properties of interest, in this case the C~{\sc iv} blueshift and $\Delta(g-i)$, 
are thus more reliably isolated for comparison with the X-ray emission. 

\acknowledgements

We acknowledge the support of {\em Chandra} X-ray center grants
GO1--2105X (SCG, WNB) and GO3--4144A (SCG, GTR, WNB).  WNB thanks NASA
LTSA grant NAG5--13035.

\end{document}